\documentclass[aps,prl,superscriptaddress,twocolumn,showpacs,floatfix]{revtex4}
\usepackage{epsfig}

\newcommand{\ie}{{\it i.e.\ }}
\newcommand{\ignore}[1]{}
\newcommand{\bx}{{\bf x}}
\newcommand{\by}{{\bf y}}

\renewcommand{\d}{{\rm d}}
\begin{document}

\title{Robust gene regulation: \\
Deterministic dynamics from asynchronous networks with delay}

\author{Konstantin Klemm}
\author{Stefan Bornholdt}
\affiliation{Interdisciplinary Center for Bioinformatics, University of Leipzig\\
Kreuzstr.\ 7b, D-04103 Leipzig, Germany
}
\date{\today}

\begin{abstract}
We compare asynchronous vs.\ synchronous update of discrete dynamical networks 
and find that a simple time delay in the nodes may induce a reproducible
deterministic dynamics even in the case of asynchronous update in random order. 
In particular we observe that the dynamics under synchronous parallel update 
can be reproduced accurately under random asynchronous serial update for a 
large class of networks. This mechanism points at a possible general principle  
of how computation in gene regulation networks can be kept in a 
quasi-deterministic ``clockwork mode'' in spite of the absence of a central clock. 
A delay similar to the one occurring in gene regulation causes synchronization in the model. 
Stability under asynchronous dynamics disfavors topologies containing loops, 
comparing well with the observed strong suppression of loops in biological regulatory 
networks. 
\end{abstract} 
\pacs{87.16.Yc,05.45.Xt,89.75.Hc,05.65.+b} 
\maketitle 
 
Erwin Schr\"odinger in his lecture ``What is life?'' held in 1943 \cite{Schroedinger} 
was one of the first to notice that the information processing performed in the 
living cell has to be extremely robust and therefore requires a quasi-deterministic 
dynamics (which he called ``clockwork mode''). The discovery of a ``digital''  
storage medium for the genetic information, the double-stranded DNA, 
confirmed one important part of this picture. 
Today, new experimental techniques allow to observe the dynamics of 
regulatory genes in great detail, which motivates us to reconsider the other, 
dynamical part of Schr\"odinger's picture of a ``clockwork mode''.  
While the dynamical elements of gene regulation often are known in great detail, 
the complex dynamical patterns of the vast network of interacting regulatory genes,
while highly reproducible between identical cells and organisms under similar 
conditions, are largely not understood. Most remarkably, these virtually 
deterministic activation patterns are often generated by asynchronous genetic 
switches without any central clock. In this Letter we address this astonishing fact 
with a toy model of gene regulation and study the conditions of when deterministic 
dynamics could occur in asynchronous circuits. 
Let us start from the observed dynamics of small circuits of regulatory genes, 
then derive a discrete dynamical model gene, followed by a study of 
networks of such genetic switches, with a focus on comparing their 
asynchronous and synchronous dynamics. 

Recently, several small gene regulation circuits have been described in terms of 
a detailed picture of their dynamics \cite{Elowitz,Hes1,Baltimore,p53,Smolen}.  
A particularly simple motif is the single, self-regulating gene \cite{Rosenfeld,Hes1}  
that allows for a detailed modeling of its dynamics.  
A set of two differential equations, for the temporal evolution 
of the concentrations of messenger RNA and protein, respectively, 
and an explicit time delay for transmission delay provide a quantitative 
model for the observed dynamics in this minimal circuit \cite{Jensen03}. 
The equations of this model take the basic form 
\begin{eqnarray} \label{eq:originaldiff1}
\frac {\d c}{\d t} &=& \alpha [f(s(t-\vartheta)) - c(t)] \\
\frac {\d b}{\d t} &=& \beta [c(t)-b(t)]
\end{eqnarray}
for the the dynamics of the concentrations $c$ of mRNA and  $b$ of protein, 
with some non-linear transmission function $f(s)$ of an input signal $s$, 
a time delay $\vartheta$, and the time constants $\alpha$ and $\beta$. 
In order to define a minimal discrete gene model let us keep the basic 
features (delay, low pass filter characteristics), omit the second filter, 
and write the difference equation for one gene $i$ as
\begin{equation}
\Delta c_i = \alpha [f(s_i(t-\vartheta)) - c_i(t)] \Delta t~.
\end{equation}
The non-linear function $f$ is typically a steep sigmoid. We approximate it
as a step function $\Theta$ with $\Theta(s)=0$ for $s<0$ and $\Theta(s)=1$
otherwise. Rescaling time with $\epsilon = \alpha \Delta t$ and 
$\tau = \vartheta / \Delta t$ this reads 
\begin{equation}
\Delta c_i = \epsilon [\Theta(s_i(t-\tau)) - c_i(t)]~.
\end{equation}
For simplicity let us update $c_i$ by equidistant steps according to
\begin{equation} \label{eq:cupdatefinal}
\Delta c_i = \left\{ 
\begin{array}{rl}
+\epsilon, & {\rm if }\;s_i(t-\tau) \ge 0\;  {\rm and}\;c_i \le 1-\epsilon \\
-\epsilon, & {\rm if }\;s_i(t-\tau)  < 0 \;  {\rm and}\;c_i \ge   \epsilon \\
0,         & {\rm otherwise}
\end{array}
\right.
\end{equation}
The coupling between nodes is defined by
\begin{equation} \label{eq:xsum}
s_i (t) = \sum_j w_{ij} x_j (t) - a_i~,
\end{equation}
with discrete output states $x_j (t)$ of the nodes defined as
\begin{equation} \label{eq:xdefinition}
x_j (t) = \Theta(c_j (t) - 1/2)~.
\end{equation}
The influence of node $j$ on node $i$ can be activating ($w_{ij}=1$),
inhibitory ($w_{ij}=-1$), or absent ($w_{ij}=0$). 
A constant bias $a_i$ is assigned to each node.

In the following let us consider a network model of such nodes. Consider $N$ nodes 
with concentration variables $c_i$, state variables $x_i$, biases $a_i$ and a coupling 
matrix $(w_{ij})$. Given initial values $x_i(0)=c_i(0)\in\{0,1\}$ the time-discrete 
dynamics is obtained by iterating the following update steps: 

(1) Choose a node $i$ at random. 
(2) Calculate $s_i$ according to Eq.\ (\ref{eq:xsum}). 
(3) Update $c_i$ according to Eq.\ (\ref{eq:cupdatefinal}). 

For $\tau=0$ and $\epsilon=1$ random asynchronous update is recovered. 
For $\tau>0$ there is an explicit transmission delay from the output of node $j$ 
to the input of node $i$. To be definite, at $t=0$ we assume that nodes have not 
flipped during the previous $\tau$ time steps. 

Let us first explore the dynamics of a simple but non-trivial interaction network 
with $N=3$ sites and 
\begin{figure}[hbt]
\centerline{\epsfig{file=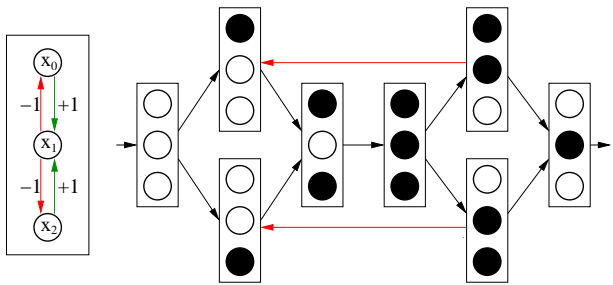,width=.45\textwidth}}
\caption{\label{fig:combined}
Left: Network with $N=3$ nodes and bias values $a_0=a_2=0$ and $a_1=2$.
Right: Dynamics of the network. Transitions between configurations under
asynchronous update are indicated by arrows. Under synchronous (parallel)
update the system has one unique cyclic attractor only, consisting of the four 
configurations in the middle row.}
\end{figure}
non-vanishing couplings $w_{01} = w_{21} = -1$ and $w_{10}= w_{12} = +1$, 
see Fig.~\ref{fig:combined}. Note that under {\em asynchronous}
update the sequence of states reached by the dynamics is not unique.
The system may branch off to different configurations depending on
node update ordering. This is illustrated in Fig.~\ref{fig:singletsz_0}(a): 
Without delay ($\tau=0$) and filter ($\epsilon=1$) the dynamics is irregular,
 \ie non-periodic. 
\begin{figure}[hbt]
\centerline{\epsfig{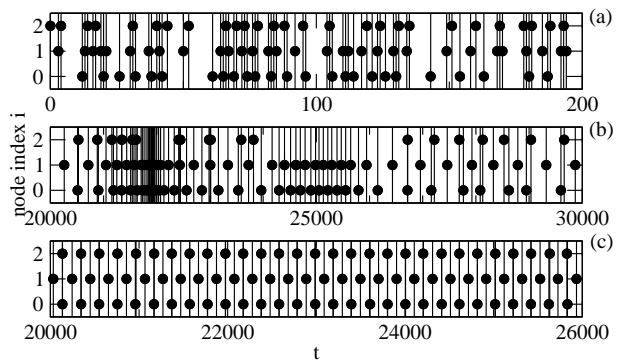}}
\caption{\label{fig:singletsz_0}
Simulation runs of the three-node network in Fig.~\ref{fig:combined}. (a)
Random asynchronous update mode, $\tau=0$, $\epsilon=1$. (b) Filtering
$\epsilon=0.01$ but no delay $\tau=0$. (c) Delay $\tau=100$ and no filtering, 
$\epsilon=1$. A circle plotted at coordinates $(t,i)$ indicates that state $x_i$ 
of node $i$ changes at time $t$.}
\end{figure}
With filter only ($\tau=0$, $\epsilon=0.01$, Fig.~\ref{fig:singletsz_0}(b)), 
the dynamics is periodic at times, but also intervals of fast irregular flipping occur. 
Finally, in the presence of delay ($\tau=100$, $\epsilon=1$, Fig.~\ref{fig:singletsz_0}(c)) 
we obtain perfectly ordered dynamics with synchronization of flips. Nodes 0 and 2 
change states practically at the same (macro) time, followed by a longer pause until 
node 1 changes state, etc. With increasing delay time $\tau$ the dynamics under 
asynchronous update approaches the dynamics under synchronous update 
(cf.~Fig.~\ref{fig:combined}) when viewed on a coarse-grained (macro) time scale.

Let us further quantify the difference between synchronous and asynchronous 
dynamics. First, a definition of equivalence between the two dynamical modes 
has to be given. Let us start from the time series $\bx (t)$ of configurations 
$\bx = (x_0,\dots, x_{N-1})$ produced by the asynchronous (random serial) 
update of the model and the respective time series $\by (u)$ produced by 
synchronous (parallel) update, using identical initial condition $\by (0) = \bx (0)$. 
These time series live on different time scales, which we call the micro time scale 
of single site updates in the asynchronous case, and the macro time scale where 
each time step is an entire sweep of the system. 
Assume that at time $t_u$ the asynchronous system is in state $\bx (t_u) = \by (u)$. 
In order to follow the synchronous update it has to subsequently reach the state 
$\by (u+1)$ on a shortest path in phase space. Formally, let us require that there is 
a micro time $t_{u+1}>t_u$ such that $\bx(t_{u+1})= \by(u+1)$ and each node flips
at most once in the time interval $[t_u,t_{u+1}]$. Once this is violated we say 
that an error has occured at the particular macro time step $u$.
This error allows to define a numerical measure of discrepancy between 
asynchronous and synchronous dynamics. Starting from identical initial conditions, 
the system is iterated in synchronous and asynchronous modes 
(here for $u_{\rm total} = 10^7$ macro time steps). Whenever the resulting time 
series are no longer equivalent, an error counter is incremented and the system 
reset to initial condition. The total error $E$ of the run is the number of errors 
divided by $u_{\rm total}$.

\begin{figure}[hbt]
\centerline{\epsfig{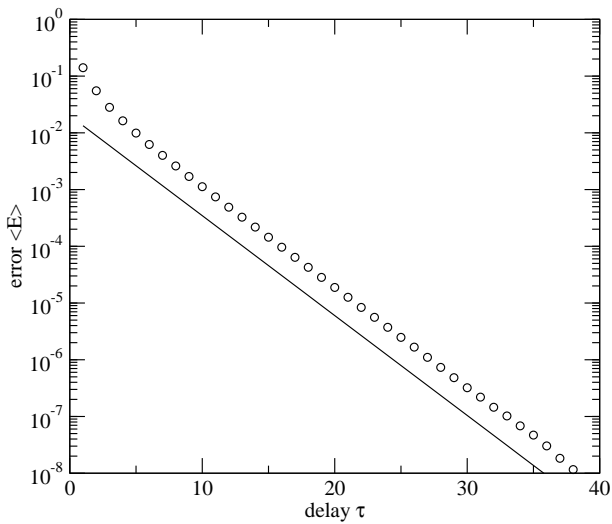}}
\caption{\label{fig:singlez_0}
Discrepancy $E$ between the asynchronous and synchronous update mode 
as a function of the delay time $\tau$ and without filter ($\epsilon = 1$). 
The solid line is the theoretical prediction of the decay $\propto (2/3)^\tau$.}
\end{figure}

For the network in Fig.\ \ref{fig:combined} and the initial condition $x_i=c_i=0$ 
for $i=1,2,3$ the error $E$ is exponentially suppressed with delay time $\tau$ 
(Fig.\ \ref{fig:singlez_0}).
The asynchronous dynamics with delay follows the attractor during a time 
span that increases exponentially with the given delay time. Note that there is only 
one possibility for the asynchronous dynamics to leave the attractor: When the
system is in configuration $(1,1,0)$ or $(0,1,1)$, node $2$ may change state
such that the system goes to configuration $(1,0,0)$ or $(0,0,1)$ respectively,
whereas the correct next configuration on the attractor is $(0,1,0)$.
Consider the case $\epsilon=1$ where $c_i=x_i$ for all $i$.
Let us assume that the system is in configuration $(1,1,1)$ and at time
$t_0$ node 0 changes state, thereby generating configuration $(0,1,1)$.
This decreases the input sum $s_1$ below zero such that for $\tau=0$
node $0$ would change state immediately in its next update. With explicit
transmission delay $\tau>0$, however, node 1 still ``sees'' the input sum $s_i=0$
generated by the configuration $(1,1,1)$ until time step $t_0+\tau$.
If node $2$ is chosen for update in this time window $t_0+1,\dots,t_0+\tau$
it changes state immediately and updates are performed in correct order.
The opposite case, that node 2 does not receive an update in any of the
$\tau$ time steps, happens with probability $(2/3)^\tau$, 
yielding the correct error decay of the simulation (Fig.\ \ref{fig:singlez_0}). 

\begin{figure}[hbt]
\centerline{\epsfig{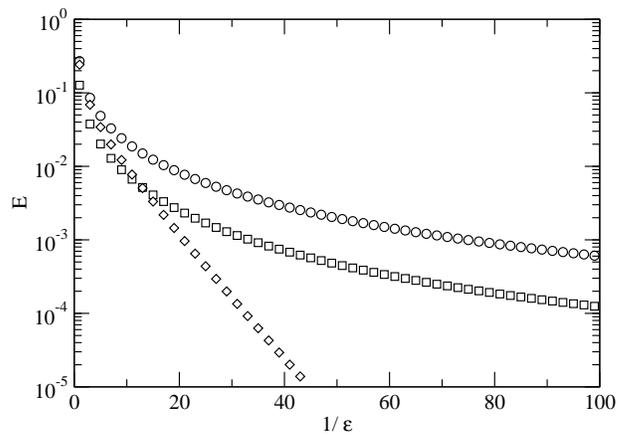}}
\caption{\label{fig:five_0}
Discrepancy $E$ between the asynchronous and synchronous update mode
as a function of the filter parameter $1/\epsilon$ for the network with
$N=5$ nodes described in the text. The delay parameter is chosen as
$\tau=0$ (circles), $\tau=10$ (squares), and $\tau=1/\epsilon$ (diamonds).}
\end{figure}

Next we demonstrate that there are cases where also low-pass filtering, 
$\epsilon \ll 1$, is needed for the asynchronous dynamics to follow the 
deterministic attractor. Consider a network of 
$N=5$ nodes with bias values $a_0 = a_4 = 0$ and $a_1 = a_2 = a_3 = 1$.
The only non-zero couplings are
$w_{10} = w_{21} = w_{31} = w_{42} = +1 $ and $w_{01} = w_{43} = -1$.
Nodes 0 and 1 form an oscillator, \ie $(x_0,x_1)$ iterate the sequence
$(0,0)$, $(1,0)$, $(1,1)$, $(0,1)$. Nodes $2$ and $3$ simply ``copy'' the
state of node $1$ such that under synchronous update always
$x_3(t)=x_2(t)=x_1(t-1)$.
Consequently, under synchronous update the input sum of node $4$ never
changes because the positive contribution from node 2 and the negative
contribution from node 3 cancel out. Under asynchronous update, however,
the input sum of node 4 may fluctuate because nodes 2 and 3 do not flip
precisely at the same time. The effect of the low-pass filter $\epsilon \ll 1$ 
is to suppress the spreading of such fluctuations on the micro time scale.
The influence of the filter is seen in Fig.~\ref{fig:five_0}. 
When $\tau$ is kept constant, the error drops algebraically with decreasing 
$\epsilon$. An exponential decay $E \sim \exp(-\alpha /\epsilon)$ is obtained
 when $\tau \propto 1/\epsilon$ (the filter can take full effect only in the
presence of sufficient delay). 

\begin{figure}[hbt]
\centerline{\epsfig{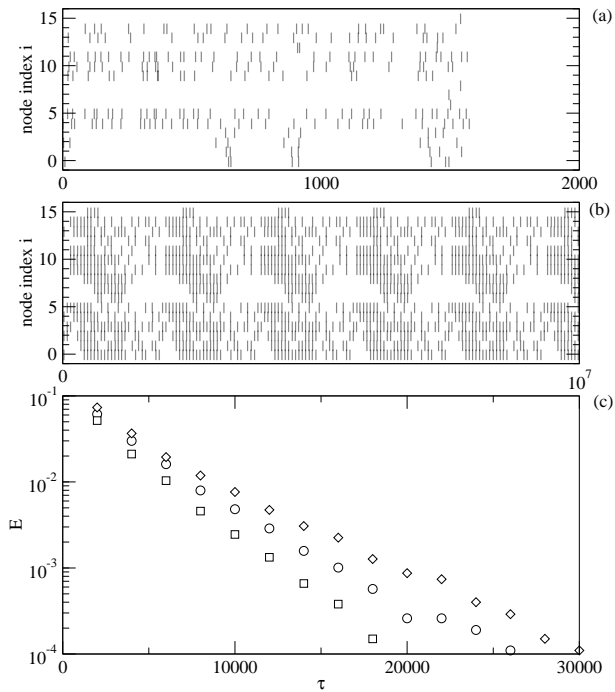}}
\caption{\label{fig:largets_1}
Time series and errors obtained for the network of $N=16$ nodes described in
the text. (a) Asynchronous update mode with neither delay nor filtering,
$\tau=0$, $\epsilon=1$. A vertical stroke at coordinates $(t,i)$ indicates that
node $i$ flips at time $t$. At $t\approx 1600$ the system reaches a fixed
point. (b) Same initial condition as in (a), but delay $\tau=50000$ and
filtering $\epsilon = 1/500$. The system follows a limit cycle of 28 macro time
steps. (c) Discrepancy (error $E$) between asynchronous and synchronous update
mode as a function of the delay parameter for $\epsilon = 50 / \tau$
(circles), $\epsilon = 40 / \tau$ (squares), $\epsilon = 25 / \tau$ (diamonds).}
\end{figure}

Let us finally consider an example of a larger network with $N=16$ nodes 
and $L=48$ non-vanishing couplings (chosen randomly from the off-diagonal 
elements in the matrix $(w_{ij})$ and assigned values $+1$ or $-1$ with 
probability $1/2$ each; biases are chosen as $a_i = \sum_j w_{ij}/2$). 
Simulation runs under pure asynchronous update ($\tau=0$, $\epsilon=1$) 
typically yield dynamics as in Fig.~\ref{fig:largets_1}(a). 
The time series ${\bx (t)}$ is non-periodic
and non-reproducible, \ie under different order of updates a different
series is obtained. For the same initial condition, periodic dynamics is
observed in the presence of sufficent transmission delay and filtering,
Fig.~\ref{fig:largets_1}(b). In this case, the system follows precisely the
attractor of period 28 found under synchronous update.
As seen in Fig.~\ref{fig:largets_1}(c), the error decays exponentially as a
function of the delay time $\tau$. 

\begin{figure}[hbt]
\vspace*{5mm}
\centerline{\epsfig{file=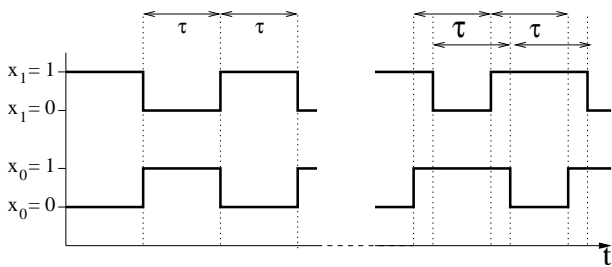,width=.45\textwidth}}
\caption{\label{fig:cycle_illu} Dynamics of the network with
$N=2$ mutually activating nodes.  Rising edges and falling
edges desynchronize over time. The two classes of edges form
two uncoupled chains of events.}
\vspace*{-5mm}
\end{figure}

Let us now turn to the dangers of asynchronous update: There is 
a fraction of attractors observed under synchronous update that 
cannot be realized under asynchronous update. Synchronization
cannot be sustained if the dynamics is separable. In the
trivial case, separability means that the set of nodes can be
divided into two subsets that do not interact with each other.
Then there is no signal to synchronize one set of nodes with
the other and they will go out of phase. In general, synchronization 
is impossible if the set of flips itself is separable.
Consider, as the simplest example, a network of $N=2$ nodes
with the couplings $w_{01} = w_{10} = +1$, biases $a_0 = a_1 =
1$ and the initial condition $(y_0(0),y_1(0)) = (0,1)$. Under
synchronous update, the state alternates between vector $(0,1)$
and $(1,0)$.  Under asynchronous update with delay time $\tau$,
the transition of one node $i$ from $x_i=0$ to $x_i=1$ causes
the other node $j$ to switch from $x_j=0$ to $x_j=1$
approximately $\tau$ time steps later. The ``on''-transitions
only trigger subsequent ``on''-transitions and, analogously,
the ``off''-transitions only trigger subsequent ``off''-transitions.   
The dynamics can be divided into two distinct sets of events that 
do not influence each other. Consequently, synchronization 
between flips cannot be sustained, as illustrated 
in Fig.\ \ref{fig:cycle_illu}. 
When the phase difference reaches the value $\tau$, on- and
off-transitions annihilate. Then the system leaves the
attractor and reaches one of the fixed points with $x_0 = x_1$. 

These observations have important implications for robust  
topological motifs in asynchronous networks. First of all, 
the above example of a small excitatory loop can be quickly 
generalized to any larger loop with excitatory interactions, 
as well as to loops with an even number of inhibitory couplings, 
where in principle similar dynamics could occur. Higher order
structures that fail to synchronize include competing modules, 
e.g.\ two oscillators (loops with odd number of inhibitory links) 
that interact with a common target. 

In conclusion we find that asynchronously updated 
networks of autonomous dynamical nodes are able to 
exhibit a reproducible and quasi-deterministic dynamics 
under broad conditions if the nodes have transmission 
delay and low pass filtering as, e.g., observed in regulatory genes. 
Timing requirements put constraints on the topology of 
the networks (e.g.\ suppression of certain loop motifs). 
With respect to biological gene regulation networks 
where indeed strong suppression of loop structures is observed
\cite{Shen-Orr02,Milo02}, one may thus speculate about a 
new constraint on topological motifs of gene regulation:
The requirement for deterministic dynamics from 
asynchronous dynamical networks. 

{\bf Acknowledgements}
S.B.\ thanks D.\ Chklovskii, M.H.\ Jensen, S.\ Maslov, and K.\ Sneppen for 
discussions and comments, and the Aspen Center for Physics for 
hospitality where part of this work has been done.


\begin{thebibliography}{99}

\bibitem{Schroedinger}
E.\ Schr\"odinger, {\sl What is Life? The Physical Aspect of the Living Cell}, 
Cambridge: University Press (1948). 

\bibitem{Hes1}
H.\ Hirata et al., Science {\bf 298}, 840 (2002). 

\bibitem{Elowitz}
M.B.\ Elowitz and S.\ Leibler, Nature {\bf 403}, 335 (2002). 

\bibitem{Baltimore}
A.\ Hoffmann, A.\ Levchenko, M.L.\ Scott, and D.\ Baltimore,     
Science {\bf 298}, 1241-1245 (2002). 

\bibitem{p53}
G.\ Tiana, M.H.\ Jensen, and K.\ Sneppen, 
Eur.\ Phys.\ J.\ B {\bf 29}, 135-140 (2002). 

\bibitem{Smolen}
P. Smolen, D. A. Baxter, J. H. Byrne, 
Bull. Math. Biol. {\bf 62}, 247 (2000). 

\bibitem{Rosenfeld}
N. Rosenfeld, M. B. Elowitz, U. Alon, 
J. Mol. Biol. {\bf 323}, 785 (2002). 

\bibitem{Jensen03}
M.H.\ Jensen, K.\ Sneppen, G.\ Tiana, FEBS Letters {\bf 541}, 176-177 (2003). 

\bibitem{Shen-Orr02}
S.S.\ Shen-Orr, R.\ Milo, S.\ Mangan, and U.\ Alon, Nature Genetics {31}, 64-68 (2002). 

\bibitem{Milo02}
R.\ Milo, S.\ Shen-Orr, S.\ Itzkovitz, N.\ Kashtan, D.\ Chklovskii, and U.\ Alon,
Science {\bf 298}, 824-827 (2002).

\end{thebibliography}
\end{document}